# Barrier Height Formation in Organic Blends / Metal Interfaces: Case of (TTF-TCNQ) / Au(111)

José I. Martínez[1,2,*], Enrique Abad[3], Juan I. Beltrán[2], Fernando Flores[2], and José Ortega[2]

[1] Depto. Superficies y Recubrimientos,
Instituto de Ciencia de Materiales de Madrid (CSIC), ES-28049 Madrid, Spain

[2] Depto. Física Teórica de la Materia Condensada,
Universidad Autónoma de Madrid, ES-28049 Madrid, Spain

[3] Institute of Theoretical Chemistry, Universität Stuttgart, D-70569 Stuttgart, Germany

[*] joseignacio.martinez@icmm.csic.es

## Abstract

The interface between the tetrathiafulvalene / tetracyanoquinodimethane (TTF-TCNQ) organic blend and the Au(111) metal surface is analyzed by Density Functional Theory calculations, including the effect of the charging energies on the molecule transport gaps. Given the strong donor and acceptor characters of the TTF and TCNQ molecules, respectively, there is a strong intermolecular interaction, with a relatively high charge transfer between the two organic materials. We find that the TCNQ LUMO peak is very close to the Fermi level; due to the interaction with the metal surface, the organic blend molecular levels are broadened, creating an important induced density of interface states. We show that the interface energy level alignment is controlled by the charge transfer between TTF, TCNQ and Au, and by the molecular dipoles created in the molecules because of their deformations when adsorbed on Au(111); in particular the TCNQ molecules present a bent adsorption geometry with the N atoms bonded to the Au surface. A generalization of the Unified-IDIS model, to explain how the interface energy levels alignment is achieved for the case of this blended organic layer, is presented by introducing matrix equations associated with the Charge Neutrality Levels of both organic materials and with their intermixed screening properties.

## I. INTRODUCTION

In the last years, the field of organic electronics has received a lot of attention from research and industry because of its potential use in electronic devices.[1-3] For the Surface Science Community, the study of the conductivity of these systems is very challenging,[2,4] these properties depending crucially on the different interface barriers formed at organic / organic or at metal / organic contacts.[3,5,6] For the electronic industry, these materials provide a cheap and environment-friendly way to obtain electronic devices as an alternative to conventional semiconductors. However, these materials have the drawback that their electron mobility is very low, due to the molecular character of the crystals and its weak intermolecular interaction.

The analysis of different metal / homogeneous organic interfaces has been deeply studied by a wide number of theoretical and experimental groups (e.g. see Refs. 7-17). As a result of this work, it seems fair to conclude that the organic / metal energy level alignment at the contact is the result of different mechanisms operating at the interface: charge transfer between the materials; compression of the metal electron tails due to Pauli repulsion (so called "pillow" effect); orientation of molecular dipoles; and, for reactive interfaces, formation of gap states in the organic materials. For non-reactive interfaces, the first three mechanisms (charge transfer, "pillow" effect and molecular dipoles) can be described together by means of a Unified-IDIS model[17] based on the concept of the Charge Neutrality level ($CNL$), an interface screening parameter ($S$) and the "pillow" and molecular dipoles.

One advantage of using organic semiconductors is the possibility of combining different organic materials to control and adjust the electronic properties of the barrier heights at metal / organic interfaces. These organic blends / metal interfaces have recently been explored by different groups. We mention here the work on (TTF-TCNQ) / Au(111) interfaces,[18,19,20] as well as the work by A. El-Sayed and J.E. Ortega on CuPc-PFP / Au(111) and Cu(111), and $F_{16}$CuPc-PEN / Au(111) and Cu(111).[21]

In this paper we address theoretically this problem by studying the interesting heterogeneous (TTF-TCNQ) / Au(111) interface. This interface combines two organic molecules with strongly acceptor (TCNQ) and donor (TTF) characters, this fact highlighting the possible effects associated with the difference between the electro-negativities of the organic molecules. The TTF-TCNQ crystal has been widely studied because it was the first organic conductor;[4] it has also been suggested as a good candidate for organic electronic devices due to its high mobility. The interface between the TTF and TCNQ organic crystals has been analyzed recently;[22,23] these studies show that the interface is metallic[22] due to the overlap (and charge transfer) between the density of states corresponding to the TTF highest occupied molecular orbital (HOMO) and the TCNQ lowest unoccupied molecular orbital (LUMO) levels.[23]

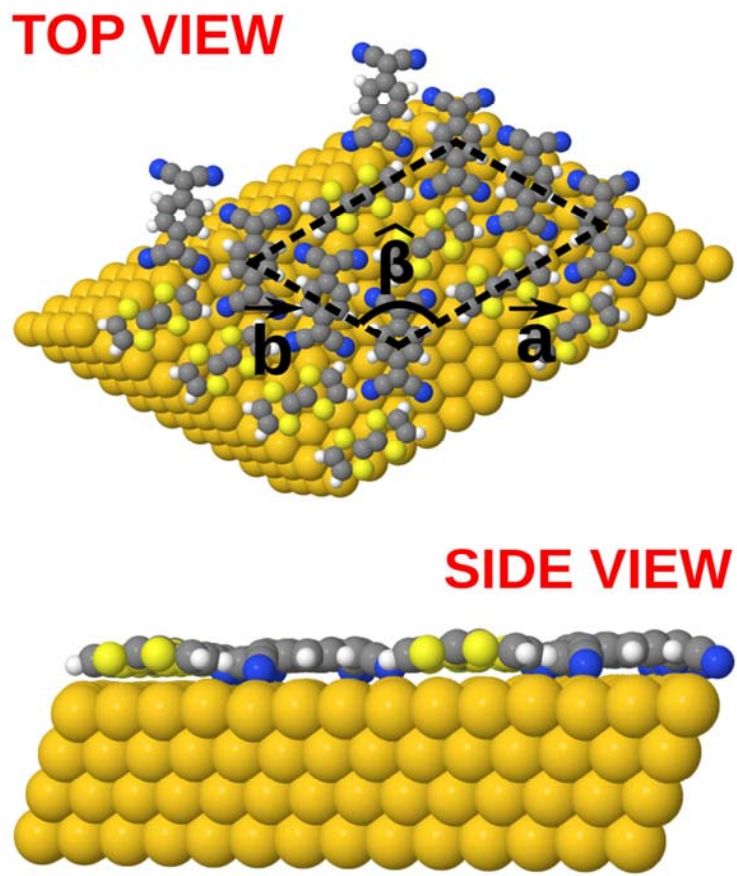

**FIG. 1** (Color online) Top and side "ball and stick" model views for the relaxed (TTF-TCNQ) / Au(111) geometry. Black dashed lines in top panel indicate the unit cell used in all the calculations.

In our approach, we have used first a local orbital DFT-method (Section II) to analyze the (TTF-TCNQ)/Au(111) interface geometry. The interface electronic structure is analyzed in Section III including in our DFT calculation the effects associated with the molecule charging energy that will allow us to correct appropriately the transport energy gap of both organic materials.[24-29] In Section IV we present a generalization of the Unified-IDIS model for the interface between a blended organic layer and a metal, introducing some matrices for the organic blends that allow us to extend to this case the concepts used previously for a homogeneous organic layer. Finally, in Section V we present our conclusions.

## II. ATOMIC STRUCTURE: LDA CALCULATIONS

In a first step, we analyze the atomic structure of the (TTF-TCNQ)/Au(111) interface. For this purpose we use the efficient local-orbital DFT code FIREBALL.[30-32] In these calculations we have used a Local Density Approximation (LDA) functional[32] and a basis set of $sp^3d^5$ numerical atomic orbitals (NAOs)[33] for Au, C, N, and S, and s for H, with cut-off radii (in a.u.): s = 4.5, p = 4.9 and d = 4.3 (Au); s = 4.0, p = 4.5 and d = 5.4 (C); s = 3.6, p = 4.1 and d = 5.2 (N); s = 4.2, p = 4.7 and d = 5.5 (S); and s=4.1 (H). This is the same basis set as used in our previous works involving the TTF/Au(111)[29] and TCNQ/Au(111)[28] interfaces. Ion-electron interaction has been modeled by means of norm-conserving scalar-relativistic pseudopotentials.[34]

In this work we have analyzed the interface geometry suggested by González-Lakunza *et al.*;[18] in this geometry the TTF and TCNQ molecules lie parallel to the Au(111) surface, forming alternating rows of TTF and TCNQ molecules. In our calculations we have initially placed the molecules on the Au(111) surface following this pattern, and then the atomic geometry has been fully relaxed by using the FIREBALL code. Lattice vectors and unit cell for the periodic DFT calculation are shown (dashed black line) in top panel of Fig. 1 (**a**=20.49 Å; **b**=14.64 Å; $\beta$ = 120º). The Au(111) surface is represented by a slab with 4 Au layers, and there are 2 TTF and 2 TCNQ molecules per unit cell. In total, the unit cell contains 208 atoms. The Brillouin zone (BZ) has been sampled by means of a [2×4×1] Monkhorst-Pack grid,[35] guaranteeing a full convergence in energy and density.

The relaxed interface geometry is shown in Figure 1. The final adsorption geometry for the TCNQ molecule is very similar to the one found in a previous work for the TCNQ/Au(111) interface; in particular, we find that the N atoms bond to the Au surface, resulting in a bent geometry for the adsorbed TCNQ molecules.[28] The adsorption distance, however, of TCNQ on the surface increases up to a value of 2.96 Å (to be compared with that of 2.85 Å in its homogeneous interface). On the other hand, the adsorption distance of TTF on the surface decreases up to a value of 2.83 Å (3.05 Å in its homogeneous interface [29]).

## III. ELECTRONIC STRUCTURE: CHARGING ENERGIES AND INTERFACE ENERGY LEVEL ALIGNMENT

In a second step we analyze the electronic structure and energy level alignment of the (TTF-TCNQ)/Au(111) interface. For this purpose, it is important to take into account that in standard DFT calculations the Kohn-Sham energy levels do not properly describe the electronic energy levels of the system and transport gaps are usually too small.[3] For example, the experimental gap between the ionization and the affinity levels of the gas-phase TCNQ (TTF) molecule is around 5.3 (6.3) eV, see Figure 2 (left panel), while the energy gap between the Kohn-Sham HOMO and LUMO levels in LDA or generalized gradient approximation (GGA) calculations are 1.6 (2.6) eV.[23]

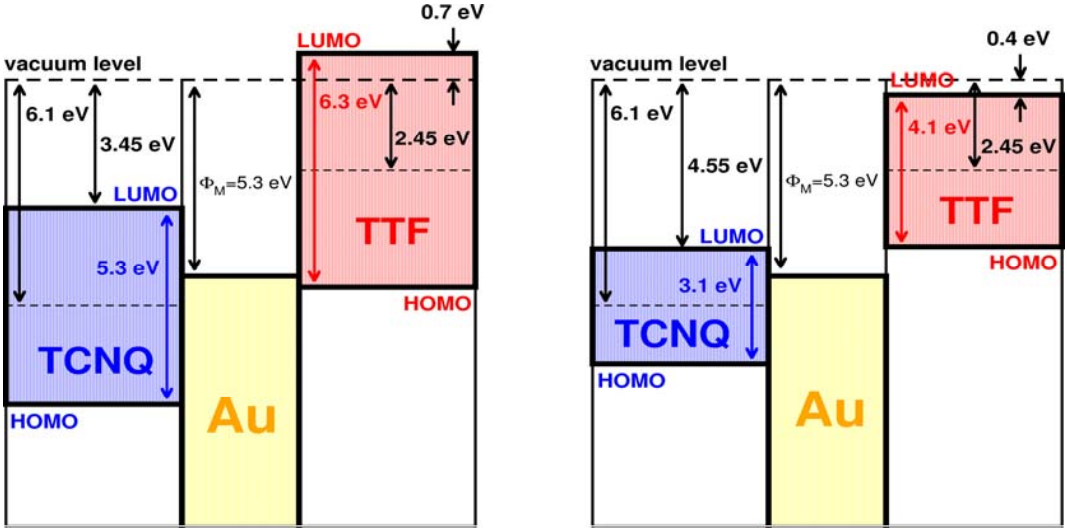

**FIG. 2** (Color online) Energy levels positions for the gas-phase (left panel) TTF and TCNQ molecules with respect to the Au(111) surface. The right panel shows how dynamical polarization or screening effects at the interface reduce the transport gaps for TTF and TCNQ, shifting in different directions occupied and empty states.[3]

In our calculations for the (TTF-TCNQ)/Au(111) interface we have corrected for this deficiency by introducing for each molecule a scissor operator:

$$O_\alpha^{scissor} = \frac{U_\alpha}{2}\sum_{(\mu\nu)}\{|\mu_i\rangle\langle\mu_i| - |v_i\rangle\langle v_i|\}, \quad (1)$$

$|\mu_i\rangle$ and $|v_i\rangle$ being the empty (occupied) orbitals of the isolated molecule (with the actual geometry of the molecule on the surface); $U_\alpha$ is the charging energy of the α-molecule. In this work we take the values of $U_\alpha$ as calculated in Refs. 28, 29 for the TCNQ/Au(111) and TTF/Au(111) interfaces; this yields the following transport energy gaps for TTF and TCNQ on the Au(111) surface: $E^t(TTF)$ = 4.1 eV and $E^t(TCNQ)$ = 3.1 eV (see Fig. 2, right panel).

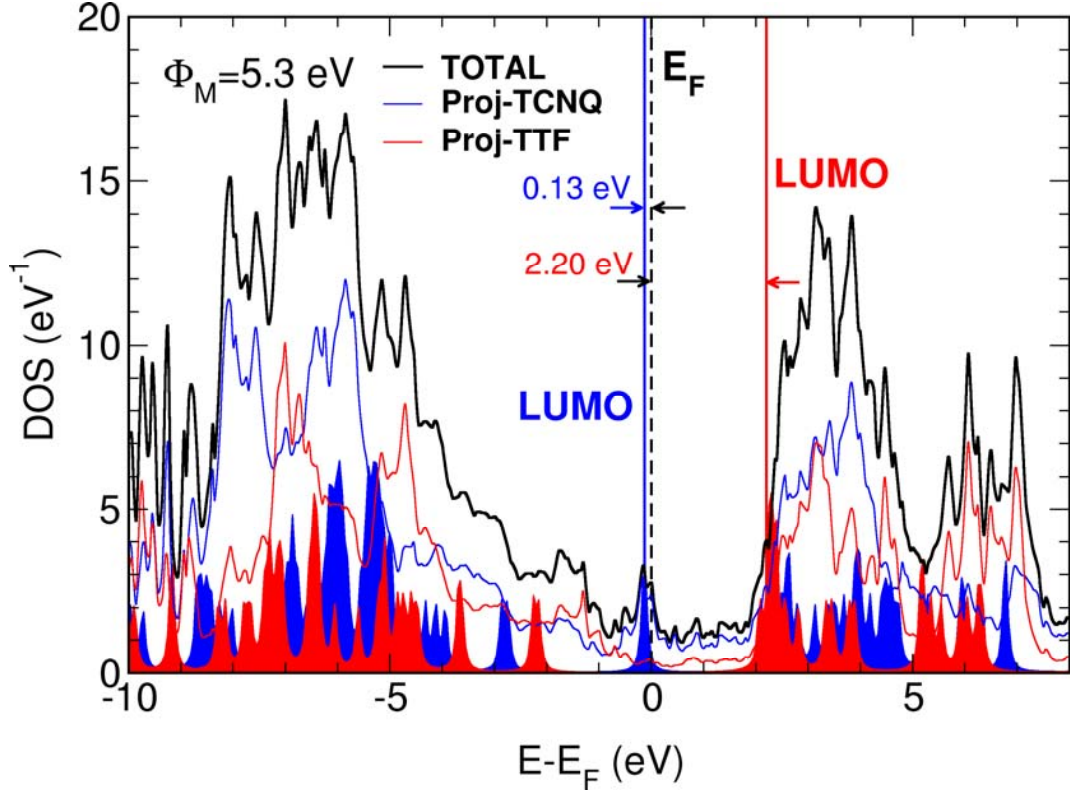

**FIG. 3** (Color online) Projected Density of States (in eV$^{-1}$) onto the TTF and TCNQ molecules for $\Phi_M$=5.3 eV in the (TTF-TCNQ)/Au(111) interface – referred to the Fermi energy –. (Black solid line) total DOS profile; (blue and red solid lines) TCNQ and TTF contributions, respectively; and (blue and red shaded regions) TCNQ and TTF gas-phase molecules (see text), respectively.

Figure 3 shows the electron density of states for the (TTF-TCNQ)/Au(111) interface projected onto the orbitals of the two molecules, as well as the sum of both density of states profiles. In the same figure, we also show the energy levels of the isolated (but deformed) molecules; for comparison, the transport gaps of these isolated molecules have been corrected using the same values of $U_\alpha$ introduced for the calculation of the (TTF-TCNQ) / Au(111) interface. In this figure the Fermi energy, $E_F$, as well as the LUMO and HOMO levels of both molecules, are shown. Notice that the main effect of the contact is to broaden the molecular levels, creating an important induced density of interface states (clearly seen in the TTF or the TCNQ energy gaps) and a shift of the molecular levels with respect to $E_F$ due to the induced potential on each molecule.

The results shown in Fig. 3 can be compared with the experimental evidence for this interface.[18,19] The experimental spectroscopy data show that the TTF LUMO peak is 1.7 eV above $E_F$,[18] to be compared with the theoretical value of 2.2 eV above $E_F$. On the other hand, a Kondo resonance associated with the TCNQ LUMO level has been observed in this system by Fernández-Torrente et al.[19] suggesting that the TCNQ LUMO level, as calculated in our DFT approach, should be located below (and close to) the Fermi level, in good agreement with our calculations. We should also mention that other structures found in the spectroscopy of González-Lakunza et al.[18] have been shown by these authors to be associated with metal surface states resonances. Thus, we conclude that the results shown in Fig. 3 are in good agreement with the experimental evidence.[18,19]

In order to obtain an appropriate description of the interaction between the metal and the organic molecules, we have found convenient to analyze the same interface changing fictitiously the metal work-function $\Phi_M$ (=-$E_F$). This can be achieved in our calculation by introducing the following shift operator $O^{shift} = \sum_\alpha \varepsilon_0 |\alpha\rangle\langle\alpha|$, $|\alpha\rangle$ being the eigenstates of the metal. Top panel of Fig. 4 shows our results for the excess of charge, $\delta n$, found for the metal and the molecules, and bottom panel of Fig. 4 the potentials induced in TTF and TCNQ molecules w.r.t. the metal as a function of the fictitious metal work-function. For $\Phi_M$=5.3 eV (Au work-function) we find that $\delta n$(TCNQ)=0.37 electrons / molecule, $\delta n$(TTF)=-0.59 electrons / molecule, and $\delta n$(metal)=0.22 electrons/(pair of TTF and TCNQ molecules). At the same time, we find that the TCNQ (TTF) levels are shifted by 0.84 (2.70) eV with respect to the metal (this positive sign means that the molecular levels are shifted downwards in energy), in such a way that the LUMO level of TCNQ is very close to $E_F$ (initially, this LUMO is around 0.7 eV above $E_F$).

## IV. IDIS MODEL ANALYSIS

The IDIS model[3,17] has been previously used to analyze different Metal-Organic[24-29] or Organic-Organic[23] interfaces; in all these cases the organic layers are homogeneous, i.e. contain only one kind of organic molecule. In this section we extend the IDIS-model for the case of an organic layer with two different types of molecules and apply it to the analysis of the (TTF-TCNQ)/Au(111) interface. The discussion presented here also shows how this analysis can be generalized for a more complex organic layer.

In a first step, we relate the induced charges on the organic molecules, $\delta n_{TTF}$ and $\delta n_{TCNQ}$, to the total induced potentials on each molecule, $V^{tot}_{TTF}$ and $V^{tot}_{TCNQ}$, by means of the following equations:[17]

$$\delta n_{TTF} = D_{TTF}[(CNL_{TTF} - E_F) - eV^{tot}_{TTF}], \qquad (2a)$$
$$\delta n_{TCNQ} = D_{TCNQ}[(CNL_{TCNQ} - E_F) - eV^{tot}_{TCNQ}]; \qquad (2b)$$

here $CNL_{TTF}$ and $CNL_{TCNQ}$ are the charge neutrality levels of both organic materials, and $D_{TTF}$ and $D_{TCNQ}$ their corresponding average density of states around the Fermi level ($E_F = -\Phi_M$); notice that the initial value of ($CNL_i - E_F$) is reduced by the induced potential, $eV^{tot}_i$, which shifts the organic levels w.r.t. the metal Fermi energy (for $eV^{tot}_i$ positive, the corresponding organic levels are shifted in the downwards direction). Equations (2a) and (2b) can be used as linear incremental equations in the interval 1.8 eV < $|\Phi_M|$ < 5.3 eV (as suggested by Fig. 4) so that:

$$\Delta(\delta n_{TTF}) = -D_{TTF}[\Delta E_F + e\Delta V^{tot}_{TTF}], \quad (3a)$$
$$\Delta(\delta n_{TCNQ}) = -D_{TCNQ}[\Delta E_F + e\Delta V^{tot}_{TCNQ}]. \quad (3b)$$

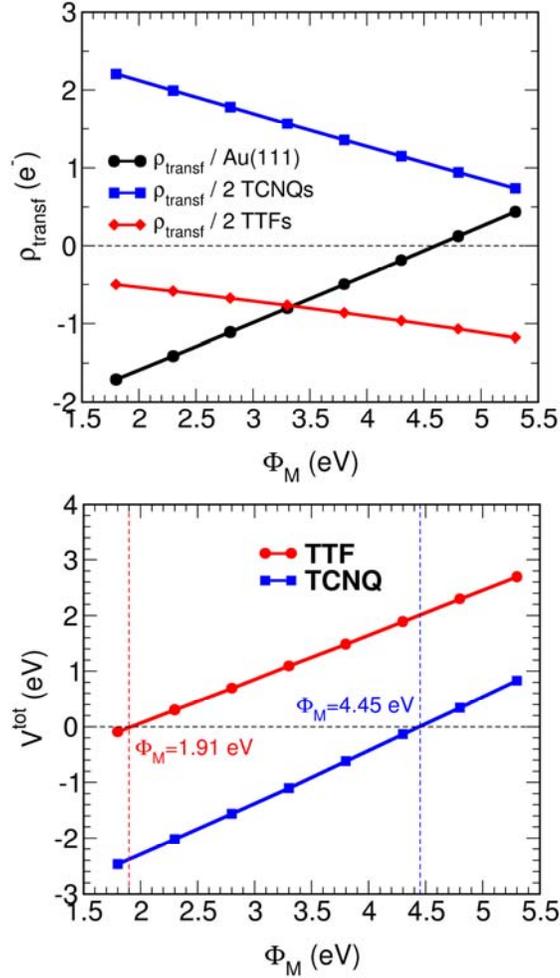

**FIG. 4** (Color online) *(Top panel)* Net charge transfer between TTF (red), TCNQ (blue) and Au(111) (black) per unit cell in the (TTF-TCNQ)/Au(111) interface as a function of a fictitious change in the metal work function; the unit cell contains 2 TTF and 2 TCNQ molecules. *(Bottom panel)* IDIS potential for the TTF and TCNQ molecules.

These equations allow us to calculate from Fig. 4 $D_{TTF}$ and $D_{TCNQ}$ since: $D_i=-\Delta(\delta n_i)/[\Delta E_F + e\Delta V^{tot}_i]$; once $D_{TTF}$ and $D_{TCNQ}$ are calculated, equations (2a) and (2b) can be used to obtain $CNL_{TTF}$ and $CNL_{TCNQ}$. This procedure yields: $D_{TTF}$=0.48 eV$^{-1}$, $D_{TCNQ}$=3.67 eV$^{-1}$, $CNL_{TTF}$=-1.2 eV and $CNL_{TCNQ}$=-4.57 eV. These values of $D_i$ are in good agreement with the ones shown in Fig. 3; notice also that for TCNQ the CNL is close to the LUMO level, while for TTF, the CNL is 0.8 eV from the LUMO (see Fig. 2).

We can also relate the potentials, $V^{tot}_{TTF}$ and $V^{tot}_{TCNQ}$, to the charges, $\delta n_{TTF}$ and $\delta n_{TCNQ}$, by the equations:

$$eV^{tot}_{TTF} = U^{eff}(TTF)\delta n_{TTF} + J^{eff}\delta n_{TCNQ} + eV^{tot(0)}_{TTF}, \qquad (4a)$$

$$eV^{tot}_{TCNQ} = U^{eff}(TCNQ)\delta n_{TCNQ} + J^{eff}\delta n_{TTF} + eV^{tot(0)}_{TCNQ}, \qquad (4b)$$

where $U^{eff}$ is a kind of charging energy for the corresponding molecule (including all the equivalent molecules of the monolayer), and $J^{eff}$ an effective interaction between the molecules.[17] $eV^{tot(0)}_i$ represents the potential induced in the i-molecule when $\delta n_{TTF}=\delta n_{TCNQ}=0$; this can be associated with the molecular intrinsic dipoles (this is important in TCNQ[28]) or the "pillow" effect[17] (which is negligible in our present case). Equations (4a) and (4b) can also be used as a linear incremental equation in the interval 1.8 < |$\Phi_M$| < 5.3 eV as follows:

$$e\Delta V^{tot}_{TTF} = U^{eff}(TTF)\Delta(\delta n_{TTF}) + J^{eff}\Delta(\delta n_{TCNQ}), \qquad (5a)$$

$$e\Delta V^{tot}_{TCNQ} = U^{eff}(TCNQ)\Delta(\delta n_{TCNQ}) + J^{eff}\Delta(\delta n_{TTF}). \qquad (5b)$$

These equations yield two constrains in the values of $U^{eff}(TTF)$, $U^{eff}(TCNQ)$ and $J^{eff}$; on the other hand, $U^{eff}(TTF)$ (or $U^{eff}(TCNQ)$) can be obtained independently by analyzing the case (not shown in this paper) of the TTF (TCNQ) layer on Au(111), eliminating the TCNQ (TTF) rows while keeping for the remaining TTF (TCNQ) rows the geometry of the full heterogeneous monolayer. This yields the following quantities: $U^{eff}(TTF)$=4.3 eV; $U^{eff}(TCNQ)$=3.7 eV; and $J^{eff}$=1.8 eV.

Coming back to equations (4a) and (4b), we can use these values to calculate $eV^{tot(0)}_{TTF}$ and $eV^{tot(0)}_{TCNQ}$: $eV^{tot(0)}_{TTF}$=0.84 eV; and $eV^{tot(0)}_{TCNQ}$=1.12 eV (notice that a positive value of $eV^{tot(0)}$ means that the molecule levels go down in energy). It is also convenient to combine equations (2a), (2b), (3a) and (3b) by introducing $\delta n_{TTF}$ – eq. (2a) – and $\delta n_{TCNQ}$ – eq. (2b) – into the equations (4a) and (4b). This yields the following matrix equations:

$$\begin{bmatrix} eV_{TTF}^{tot} \\ eV_{TCNQ}^{tot} \end{bmatrix} = (\mathbf{1} - \mathbf{S}) \begin{bmatrix} CNL_{TTF} - E_F \\ CNL_{TCNQ} - E_F \end{bmatrix} + \mathbf{S} \begin{bmatrix} eV_{TTF}^{tot(0)} \\ eV_{TCNQ}^{tot(0)} \end{bmatrix}, \quad (6a)$$

where:

$$\mathbf{S} = (\mathbf{1} + \mathbf{UD})^{-1} = \begin{bmatrix} 1 + U^{eff}(TTF)D_{TTF} & JD_{TCNQ} \\ JD_{TTF} & 1 + U^{eff}(TCNQ)D_{TCNQ} \end{bmatrix}^{-1}. \quad (6b)$$

We can also rewrite equation (6a) as follows:

$$e\mathbf{V}^{tot} = (\mathbf{1} - \mathbf{S})(\mathbf{CNL} - \mathbf{E_F} - e\mathbf{V}^{tot(0)}) + e\mathbf{V}^{tot(0)}. \quad (7)$$

Equations (6) and (7) generalize to the heterogeneous monolayer the equations for an homogeneous one; in this simple case, we have the following scalar equations:

$$eV^{tot} = (1 - S)(CNL - E_F) + SeV^{tot(0)}$$

$$= (1 - S)(CNL - E_F - eV^{tot(0)}) + eV^{tot(0)}, \quad (8)$$

where the term, *(1-S)(CNL-Φ$_M$-eV$^{tot(0)}$)* yields the induced potential associated with the charge transfer between the metal and the organic semiconductor; on the other hand, *eV$^{tot(0)}$* represents, in general, a "pillow" potential and/or the bare molecular potential associated with the molecule intrinsic dipole, with *S=1/(1+UD)* being the interface screening parameter. Notice that equations (6) and (7) can be interpreted in the same way using 2×2-matrices that incorporate the effect of both organic materials; in our particular case, *eV$^{tot(0)}$* is mainly due to the intrinsic dipole associated with TCNQ (notice that the "pillow" effect is small for both TTF and TCNQ[28,29]). It is interesting to realize that for *J$^{eff}$* =0, equations (5) decouple into the equations for two independent monolayers; this shows that, in this model, *J$^{eff}$* provides the coupling between the two organic materials. On the other hand, if we take *CNL$_{TTF}$* = *CNL$_{TCNQ}$*; *D$_{TTF}$* = *D$_{TCNQ}$*; *U$^{eff}$(TTF)* = *U$^{eff}$(TCNQ)* and *eV$^{tot(0)}$$_{TTF}$* = *eV$^{tot(0)}$$_{TCNQ}$* we also recover the monolayer case with an effective *U* given by *U$^{eff}$* + *J$^{eff}$*. In our case we obtain the following quantities: *U$^{eff}$(TTF)D$_{TTF}$* = 2.05 and *J$^{eff}$ D$_{TCNQ}$* = 6.61; and *J$^{eff}$ D$_{TTF}$* = 0.86 and *U$^{eff}$(TCNQ)D$_{TCNQ}$* = 13.6. Then,

$$\mathbf{S} = \begin{bmatrix} 0.375 & -0.17 \\ -0.022 & 0.079 \end{bmatrix}, \quad (9a)$$

$$\mathbf{1} - \mathbf{S} = \begin{bmatrix} 0.625 & 0.17 \\ 0.022 & 0.921 \end{bmatrix}. \quad (9b)$$

This yields:

$$eV_{TTF}^{tot} = 0.625(CNL_{TTF} - E_F - eV_{TTF}^{tot(0)}) +$$

$$+0.17(CNL_{TCNQ} - E_F - eV_{TCNQ}^{tot(0)}) + eV_{TTF}^{tot(0)}, \quad (10a)$$

$$eV_{TCNQ}^{tot} = 0.022(CNL_{TTF} - E_F - eV_{TTF}^{tot(0)}) +$$

$$+0.921(CNL_{TCNQ} - E_F - eV_{TCNQ}^{tot(0)}) + eV_{TCNQ}^{tot(0)}. \quad (10b)$$

These equations embody the main effects found in our DFT-based calculations. Notice first that from these equations we obtain the following relations: $\Delta V_{TTF}$ = 0.795 $\Delta \Phi_M$, and $\Delta V_{TCNQ}$ = 0.952$\Delta \Phi_M$, in good agreement with Fig. 4. On the other hand, we find, by neglecting the coupling between the two organic molecules, that:

$$eV_{TTF}^{tot} \cong 0.625(CNL_{TTF} - E_F - eV_{TTF}^{tot(0)}) + eV_{TTF}^{tot(0)}, \quad (11a)$$

$$eV_{TCNQ}^{tot} \cong 0.921(CNL_{TCNQ} - E_F - eV_{TCNQ}^{tot(0)}) + eV_{TCNQ}^{tot(0)}, \quad (11b)$$

which can be interpreted, in a first approximation, as the independent behavior of each organic sub-monolayer. The TTF-TCNQ interaction introduces the changes 0.17($CNL_{TCNQ}$ - $E_F$ -$eV^{tot(0)}{}_{TCNQ}$) and 0.022($CNL_{TTF}$- $E_F$ -$eV^{tot(0)}{}_{TTF}$) in $eV^{tot}{}_{TTF}$ and $eV^{tot}{}_{TCNQ}$, respectively.

Fig. 5 shows an energy scheme indicating how the organic energy levels are shifted with respect to the metal ($\Phi_M$=5.3 eV): in a first step, we have the bare potential shifts (due to the intrinsic molecular dipoles), $eV^{tot(0)}{}_{TTF}$ and $eV^{tot(0)}{}_{TCNQ}$ (0.84 and 1.1 eV, respectively); in a second step, the charge transfer introduces new shifts, around 1.9 eV for TTF and -0.30 eV for TCNQ. It is remarkable that for this work-function, neglecting the cross terms – 0.17($CNL_{TCNQ}$-$E_F$-$eV^{tot(0)}{}_{TCNQ}$) and 0.022($CNL_{TTF}$-$E_F$-$eV^{tot(0)}{}_{TTF}$) – introduces an error in the potential shifts smaller than 0.1 eV. However, for the fictitious case $\Phi_M$= 1.8 eV, that error would be as large as 0.6 eV for $eV^{tot}{}_{TTF}$.

## V. SUMMARY AND CONCLUSIONS

In summary, we have analyzed the energy level alignment for the interface between a metal and an organic adlayer that consists of a blend of two different organic molecules. In particular, we have studied the (TTF-TCNQ)/Au(111) interface; this interface combines two organic molecules with strong donor (TTF) and acceptor (TCNQ) characters. The experimental evidence shows that the TTF and TCNQ molecules lie parallel to the Au(111) surface, forming alternating rows of TTF and TCNQ molecules. In our analysis, we start with this adlayer geometry, as suggested in Ref. [18], and relax the atomic positions by using a local-orbital DFT calculation. The TTF and TCNQ geometries after relaxation in the heterogeneous (TTF-TCNQ)/Au(111) interface are similar to those found for their homogeneous TTF/Au(111) and TCNQ/Au(111) interfaces, although the distance between TCNQ and the Au surface is increased while the distance between TTF and Au is reduced as compared with the homogenous cases.

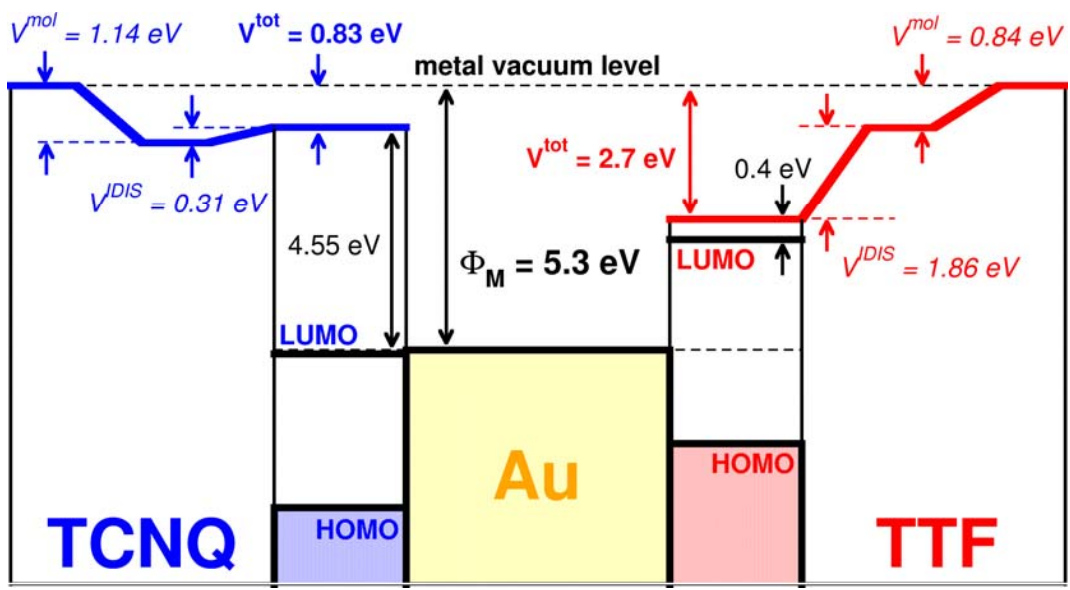

**FIG. 5** (Color online) Energy levels scheme indicating how the organic energy levels are shifted with respect to the metal ($\Phi_M$=5.3 eV) after the formation of the (TTF-TCNQ) / Au(111) interface.

The electronic structure and energy level alignment for this interface are analyzed introducing appropriate corrections to properly describe the energy gaps for the TTF and TCNQ molecules in the DFT calculation. These corrections are related to the charging energies $U_\alpha$ of the two organic molecules on the (TTF-TCNQ)/Au(111) interface; in the present calculations we have used the values for the charging energies obtained in previous calculations for the TTF/Au(111) or TCNQ/Au(111) interfaces.[28,29] These calculations show that the TCNQ LUMO peak is very close to the Fermi level. Due to the interaction with the metal surface, the molecular levels are broadened, creating a significant induced density of interface states. We find an important transfer of charge between the two organic materials, and between the organic layer and the Au(111) surface; in particular, we obtain δn(TCNQ)=0.37 electrons / (TCNQ-molecule), δn(TTF)=-0.59 electrons / (TTF-molecule) and δn(metal)=0.22 electrons/(pair of TTF and TCNQ molecules), the important transfer of charge between the TTF and TCNQ molecules reflecting their donor and acceptor characters. The induced potentials on the TCNQ and TTF molecules are 0.84 and 2.70 eV, respectively, shifting the molecular levels downwards in energy w.r.t. the metal.

The results from the DFT calculations have been analyzed by means of an extension of the Unified-IDIS model for the case of a blended organic layer. In Section IV this generalization is presented in detail and applied to the (TTF-TCNQ)/Au(111) interface. In this case we have two different types of organic molecules in the adlayer; the basic scalar equation – eq. (8) – relating the induced potential in the organic layer with the *CNL* of the organic material, the screening parameter *S* and the molecular dipole potential $eV^{tot(0)}$ is transformed into a matrix equation – eqs. (6-7) – with 2×2 matrices (e.g. for **S**) incorporating the effect of both organic materials. Finally, we mention that the IDIS formalism presented here can be naturally extended for more complex heterogeneous organic blend / metal interfaces involving a higher number of organic molecules.

**ACKNOWLEDGEMENTS**

Present work was supported by Spanish MICIIN under contract FIS2010-16046, the CAM under contract S2009/MAT-1467, and the European Project MINOTOR (grant FP7-NMP-228424). JIM acknowledges funding from Spanish MICIIN and CSIC through Juan de la Cierva and JaeDoc Programs.